\documentclass[10pt,aps,prl,twocolumn,showpacs,floatfix]{revtex4-1}

\usepackage[pdftex]{graphicx}
\usepackage{amssymb}
\usepackage{amsmath}
\usepackage{xspace}
\usepackage{color}

\begin{document} 
\title{Criticality at the Haldane-insulator charge-density-wave  quantum phase transition  }
\author{Florian Lange}
\author{Satoshi Ejima}
\author{Holger Fehske}
\affiliation{Institut f\"ur Physik, Ernst-Moritz-Arndt-Universit\"at
Greifswald, 17489 Greifswald, Germany}

\date{\today}

\begin{abstract}
Exploiting the entanglement concept within a matrix-product-state based infinite density-matrix renormalization group approach, 
we show that the spin-density-wave and bond-order-wave ground states of the one-dimensional half-filled extended Hubbard model give way to a symmetry-protected topological Haldane state in case an additional alternating ferromagnetic spin  interaction is added. In the Haldane insulator 
the lowest entanglement level features a characteristic twofold
 degeneracy. Increasing the ratio between nearest-neighbor and local
 Coulomb interaction $V/U$,  the enhancement of the entanglement
 entropy, the variation of the charge, spin and neutral gaps, and the
 dynamical spin and density response signal a quantum phase transition  to a charge-ordered state. Below a critical point, which belongs to the
universality class of the tricritical Ising model with central charge 7/10, the model is critical with $c=1/2$ along the
transition line. Above this point,  the transition between the Haldane insulator and charge-density-wave phases becomes first
order. 
\end{abstract}

\maketitle

Topological phases of matter have become one of the most fascinating objects of investigation in solid state physics~\cite{Mo10,HK10,QZ11}. 
Topological states may arise from topological band structures or interactions. The order associated with these phases can be described by topological invariants that are insensitive to gradual changes of the system parameters. As a generic feature, topological ordered states contain gapless edge excitations that encode all the information of bulk topological order~\cite{Ha82b}. 

Symmetry-protected topological (SPT) phases are zero-temperature quantum states with a given symmetry and a finite energy 
gap.  The Landau symmetry breaking states belong to this class. However, there are more interesting SPT states that do not break 
any symmetry.  For example, in higher dimensions, the Kane-Mele band insulator~\cite{KM05,KM05b} is a topological state protected by $U(1)$ 
and time-reversal symmetries. In one dimension, a prominent representative is the Haldane phase in the spin-1 Heisenberg chain~\cite{Ha83}, which 
is protected by inversion, time-reversal, and dihedral symmetries~\cite{GW09,PBTO12}. If at least one of these symmetries is not explicitly broken, 
the odd-$S$ Haldane insulator (HI) is separated from the topologically trivial state by a quantum phase transition.  Relating topological order and entanglement  allows for a further classification of topological states~\cite{CGZLW13}. While gapped quantum systems without any symmetry split in short- and  long-range entangled states, the SPT phases are always short-range entangled.   

Exploring the connection between topological band structures and interacting topological states, it has been demonstrated that the topological  invariants of gapped fermionic systems described by the one-dimensional half-filled Peierls-Hubbard model, deep in the Mott insulating regime, can be efficiently computed numerically by adding a ferromagnetic spin exchange~\cite{MENG12}. On account of a topological invariant of 2, the Peierls-Hubbard model---in a certain parameter regime---possesses the same boundary states as the spin-1 Heisenberg chain. This raises the question whether the spin-density-wave (SDW) and bond-order-wave (BOW) ground states of the half-filled extended Hubbard model (EHM)~\cite{EN07} also disappear in favor of a SPT HI phase  when a ferromagnetic spin interaction is added.  If the answer is yes, one should expect a novel quantum phase transition from the SPT state to the charge-density-wave (CDW) insulator.

In this Rapid Communications, we therefore investigate the ground-state, spectral and dynamical properties of the  EHM with additional, alternating, ferromagnetic spin coupling $J$, using the  unbiased matrix-product-state (MPS) based infinite density-matrix
renormalization group (DMRG)~technique~\cite{Wh92,Mc08,Sch11,KZMRBP13}.

The Hamiltonian of the one-dimensional EHM is 
\begin{eqnarray}
H_{\textrm{EHM}} &=&
 -t \sum_{j,\sigma}(c^\dagger_{j \sigma}c_{j+1
\sigma}^{\phantom{\dagger}} + {\rm H.c.}) 
+U \sum_j n_{j \uparrow}n_{j \downarrow} 
\nonumber\\
&&+ V \sum_{j \sigma \sigma^\prime} n_{j \sigma}n_{j+1 \sigma^\prime}. 
\end{eqnarray}
Here, $c_{j \sigma}^\dagger$ ($c_{j \sigma}^{\phantom{\dagger}}$) creates (annihilates) 
an electron with spin $\sigma$ at site $j$, $n_{j\sigma}=c_{j\sigma}^\dagger c_{j \sigma}^{\phantom{\dagger}}$, 
$t$ is the transfer amplitude of the particles, and $U$ ($V$) denotes their intrasite (intersite) Coulomb repulsion. 
We focus on the half-filled band case. 

The ground-state phase diagram of the EHM has been worked out by 
various analytical~\cite{TF02,TTC06} and numerical~\cite{Je02,SSC02,SBC04,EN07} techniques. 
In the absence of $V$ (Hubbard model), the ground state is a quantum
critical spin-density wave (SDW)
with gapless spin and gapped charge excitations $\forall$ $U>0$~\cite{EFGKK05}. 
If $2V/U\lesssim 1$, the ground state resembles that at $V=0$. 
For $2V/U\gtrsim1$, the system becomes a 2$k_{\rm F}$-CDW state, 
where both spin and charge excitation spectra are gapful. 
The SDW and CDW phases are separated by a narrow intervening BOW phase~\cite{Na99,Na2000} 
below the critical end point $[U_{\rm e},V_{\rm e}]=[9.25t, 4.76t]$~\cite{EN07}, where the ground state exhibits  
a staggered modulation of the kinetic energy density  (cf. the schematic representations included in Fig.~\ref{pd}  below).

Here we consider the extended Hamilton operator
\begin{eqnarray}
 H = H_{\textrm{EHM}}+J\sum_{j=1}^{L/2}\mbox{\boldmath
  $S$}_{2j-1}\mbox{\boldmath $S$}_{2j} 
\label{hamil}
\end{eqnarray}
with $\mbox{\boldmath $S$}_j=
   (1/2)\sum_{\sigma\sigma^\prime}
      c_{j,\sigma^{\phantom{\prime}}}^\dagger 
      \mbox{\boldmath $\sigma$}_{\sigma\sigma^\prime}^{\phantom{\dagger}}
      c_{j,\sigma^\prime}^{\phantom{\dagger}}$.
The nearest-neighbor Heisenberg spin  interaction is assumed to be alternating and ferromagnetic, i.e., $J<0$ on every other bond.   
Since the EHM at large enough $U/V$ can be thought of as  
spin-1/2 chain, the second term in \eqref{hamil} tends to form a spin-1 moment 
out of two spins on sites $2j-1$ and $2j$ in this limit. 
Then, the resulting spin-1 antiferromagnetic chain may realize a gapped Haldane
phase with zero-energy edge excitations~\cite{PBTO12}. 

To proceed  we perform an entanglement analysis of the model~\eqref{hamil}. 
The concept of entanglement is inherent in the MPS-based DMRG
algorithms too. The so-called entanglement spectrum ${\epsilon_\alpha}$ characterizes
topological phases~\cite{LH08}, which can be obtained from the singular value
decomposition. Dividing a system into two subblocks, 
${\cal H}={\cal H}_{\rm L}\otimes{\cal H}_{\rm R}$, 
and considering the reduced density matrix $\rho_{\rm L}={\rm Tr}_{\rm R}[\rho]$, $\epsilon_\alpha= -2\ln\lambda_\alpha$ are given by the singular values $\lambda_\alpha$ of the reduced density matrix $\rho_{\rm L}$. The  $\epsilon_\alpha$ spectrum also provides valuable information about the criticality of the system. Adding up the singular values $\lambda_\alpha$, we have direct access to the entanglement entropy $S_{\rm E}=-\sum_\alpha\lambda_\alpha^2\ln\lambda_\alpha^2$.
For a critical system with central charge $c$, the entanglement entropy
$S_{\rm E}$ between the two halves of the infinite chain scales 
as~\cite{CC04,PMTM09}
\begin{eqnarray}
 S_{\rm E}=\frac{c}{6}\ln\xi_\chi+s_0,
\label{se-cc}
\end{eqnarray}
where $s_0$ is a non-universal constant. The correlation length $\xi_\chi$ is determined from the second largest
eigenvalue of the transfer matrix for some bond dimension $\chi$
used in the iDMRG simulation~\cite{Mc08,Sch11,KZMRBP13}. At the critical point the physical correlation
length 
diverges, while $\xi_\chi$ stays finite due to 
the finite-entanglement cut-off. Nevertheless, $\xi_\chi$ can be used to determine the phase transition  
because it increases rapidly  with $\chi$  near the critical point. Here, we perform  iDMRG runs with $\chi$  
up to 400, so that the effective correlation length at criticality is $\xi_\chi\lesssim400$. 
\begin{figure}[t]
 \includegraphics[width=0.9\columnwidth,clip]{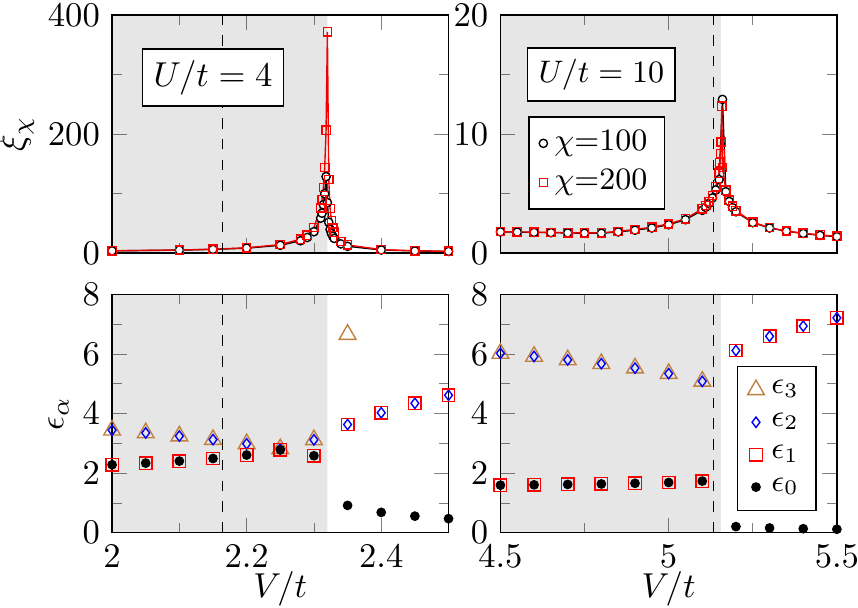}
 \caption{(Color online)  
 Correlation length $\xi_\chi$ (upper panels) and 
 entanglement spectrum $\epsilon_\alpha$ (lower panels) 
 as a function of $V/t$ for $U/t=4$ (left panels) and $U/t=10$ (right panels), where $J/t=-1.5$.   
 Data obtained by  iDMRG.  Dashed lines give the BOW-CDW (SDW-CDW)
 transition for $U/t=4$ ($U/t=10$) in the EHM~\cite{EN07}.
 }
 \label{All-u04}
\end{figure}
\begin{figure}[b]
 \includegraphics[width=0.9\columnwidth,clip]{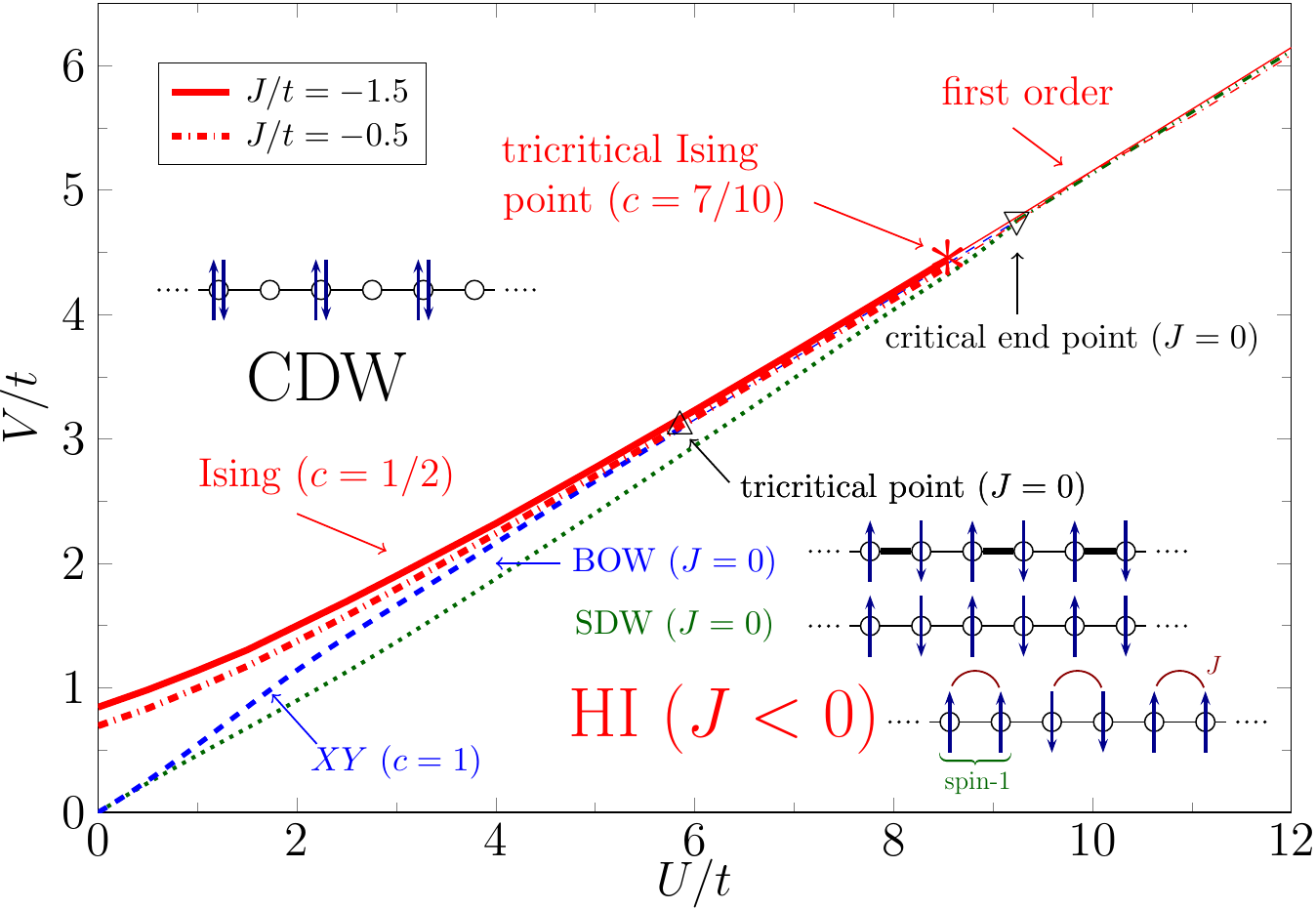} 
 \caption{(Color online) iDMRG ground-state phase diagram of  the one-dimensional (half-filled) extended
 Hubbard model with ferromagnetic spin interaction. The red solid (dotted-dashed) 
 lines give the HI-CDW phase boundaries for $J/t= -1.5$ ($-0.5$). 
 The quantum phase transition is continuous (first order) below (above) the tricritical  Ising point $[U_{\rm t}, V_{\rm t}]$ marked by 
 the star symbol.  For comparison the results for the BOW-CDW (blue dashed line), SDW-BOW (green dotted line), and SDW-CDW (green double-dotted dashed line) transitions of the pure EHM were included~\cite{EN07}.}
 \label{pd}
\end{figure}

Let us first discuss the entanglement properties of the model~\eqref{hamil}.
Figure~\ref{All-u04} shows $\xi_\chi$  and ${\epsilon_\alpha}$ in dependence on $V/t$ and $U/t$ for fixed $J/t=-1.5$.    
In the weak-to-intermediate interaction regime, $U/t=4$,  we find a pronounced peak in the correlation length at $V_{\rm c}/t\simeq2.321$,
which shoots up as $\chi$ grows from 100 to 200, indicating a divergency as $\xi_\chi\to \infty$. 
Obviously the system passes a continuous quantum phase transition. By contrast, in the strong interaction regime, $U/t=10$, the peak height 
stays almost constant when $\chi$  is raised. Decreasing $|J|$, the transition points will approach those of the pure EHM, e.g., for $J/t=-0.5$
we find $V_{\rm c}/t\simeq 2.242$, with a simultaneous reduction of the $\xi_\chi$'s peak heights. The corresponding entanglement spectra denote    that the nontrivial phase realized for $V<V_{\rm c}$  resembles the SPT Haldane phase of the spin-1 $XXZ$ model~\cite{EF15}, in that  the lowest entanglement level exhibits  a characteristic double degeneracy~\cite{PTBO10}. 
For $V > V_{\rm c}$, in the CDW phase, this level is non-degenerate.

According to Fig.~\ref{All-u04}  the maximum in the correlation length $\xi_\chi$ can be used to pinpoint the HI-CDW quantum phase transition, 
and with it map out the complete ground-state phase diagram of the EHM
with ferromagnetic spin coupling~\eqref{hamil}. The outcome is given in
Fig.~\ref{pd}, which also includes the result for the pure  EHM (blue
and green lines).  The first striking result is that the HI phase
completely replaces the SDW and BOW states. That is, the HI even
survives in the weak-coupling regime untill $U/t=0$ for any finite $J<0$
[provided that $V <V_{\rm c}(U,J)$]. According to this the itinerant
model~\eqref{hamil} behaves as a spin-1  model, even at very small
$U/t$ where double occupancy is not largely suppressed. In the intermediate-to-strong coupling regime,  the HI-CDW transition approaches the BOW/SDW-CDW transition of the EHM. The transition is continuous up to a tricritical Ising point  $[U_{\rm t}, V_{\rm t}](J)$,  which converges to the tricritical point of the  EHM as $J\to 0$. In the strong-coupling regime above $[U_{\rm t}, V_{\rm t}]$,  the HI-CDW transition becomes first order. For very large $U/t$
the phase boundaries of the HI/SDW-CDW transitions are  indistinguishable. 

\begin{figure}[t]
 \includegraphics[width=\columnwidth]{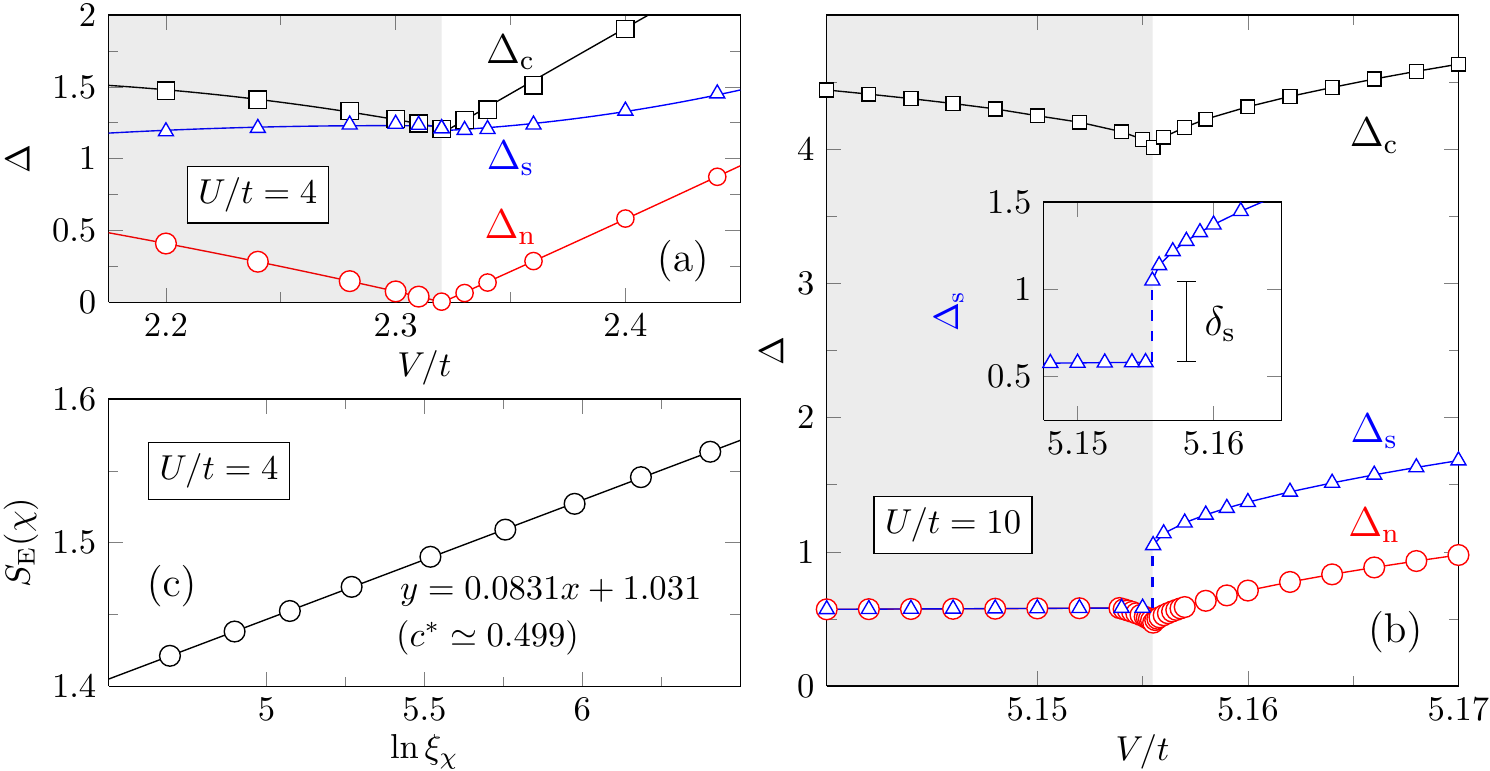}
 \caption{(Color online) 
 Charge ($\Delta_{\rm c}$), spin ($\Delta_{\rm s}$), and neutral ($\Delta_{\rm n}$) gaps as functions of $V/t$ for $U/t=4$ (a) and $U/t=10$ (b). The HI (CDW) phase is marked in grey (white).  Panel (c) gives the scaling of the entanglement entropy $S_{\rm E}(\chi)$
 with the correlation length $\xi_\chi$ at the SPT-CDW transition  $V_{\rm c}/t\simeq 2.321$ for $U/t=4$. The solid line is a linear fit of the data to Eq.~\eqref{se-cc}, indicating an Ising phase transition with $c=1/2$. 
 Results shown are obtained for $J/t=-1.5$.}
 \label{cc-gap}
\end{figure}

\begin{figure}[t]
 \includegraphics[width=.78\columnwidth]{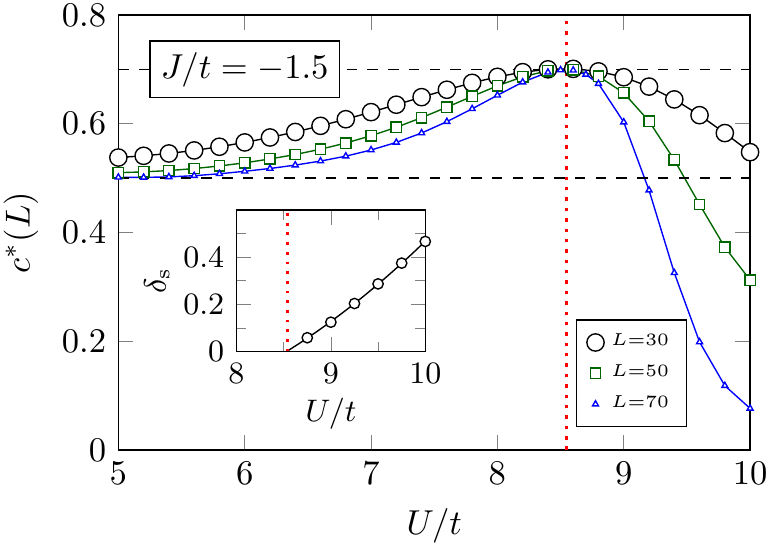}
 \caption{(Color online) Central charge $c^\ast(L)$ along the HI-CDW transition line
 for $J/t=-1.5$. DMRG data (obtained with periodic boundary conditions) indicate the  Ising universality class
 ($c=1/2$) for $U<U_{\rm t}$ and, most notably, a 
 tricritical Ising point with $c=7/10$ at $U_{\rm t}$ (red dotted line).
 Inset: Magnitude of the jump of the spin gap as $U$  further increases for $U\gtrsim U_{\rm t}$. The infinite MPS data---for a system with infinite boundary conditions---point to a first order transition. 
 }
 \label{susy}
\end{figure}

We now characterize the different states and the HI-CDW quantum phase transition in more detail.
For this we first consider the various excitation gaps:
$\Delta_{\rm c}=[E_0(N+2,0)+E_0(N-2,0)-2E_0(N,0)]/2$ [(two-particle) charge gap],
$\Delta_{\rm s}=E_0(N,1)-E_0(N,0)$ [spin gap], and $\Delta_{\rm n}=E_1(N,0)-E_0(N,0)$ [neutral gap],
where $E_0(N_{\rm e},S^z_{\rm tot})$ is the ground-state energy of the finite system with $L$ sites
for a given number of electrons $N_{\rm e}$ and $z$ component of total spin $S^z_{\rm tot}$, and 
$E_1(N_{\rm e},S^z_{\rm tot})$ is the corresponding energy of the first excited state.  
For the pure EHM, $\Delta_{\rm c}$ and  $\Delta_{\rm n}$ vanish at the BOW-CDW transition, whereas 
$\Delta_{\rm s}$ stays finite.  Here the excitation gaps were determined using DMRG in combination with 
the infinite MPS representation with `infinite boundary conditions'~\cite{PVM12,MHOV13,ZGEN12}, 
where both finite-size and boundary effects are significantly reduced.
Thereby the whole lattice  is divided into three parts: a window part, 
containing $L_{\rm W}$ sites, and two semi-infinite chains. 
While the $L_{\rm W}$-dependence persists, the $L_{\rm W}$ 
finite-size scaling is more easy to handle than the finite-size scaling 
in the traditional DMRG method. Figure~\ref{cc-gap} shows the variation
of the different excitation gaps  across the HI-CDW transition in the 
weak-coupling [Fig.~\ref{cc-gap}(a)] and strong-coupling [Fig.~\ref{cc-gap}(b)] regime. In the former case,
the charge and spin gaps feature weak minima at the transition point, but stay finite.
The neutral gap, on the other hand, closes, see Fig.~\ref{cc-gap}(a). This is evocative of the Ising transition
between the Haldane and antiferromagnetic phases in the spin-1 $XXZ$ model with single-ion 
anisotropy~\cite{EF15}. For $U/t=4$, we find $V_{\rm c}/t\simeq 2.321$. In the latter case, also the neutral gap stays 
finite passing the phase transition [see Fig.~\ref{cc-gap}(b)]. However, the jump of the spin gap 
$\delta_{\rm s}\equiv \Delta_{\rm s}(V^+_{\rm c})-\Delta_{\rm s}(V^-_{\rm c})$ 
is striking, indicating a first-order transition.
We obtain $V_{\rm c}/t\simeq5.155$ for $U/t=10$.

Next we ascertain the universality class of the HI-CDW quantum phase transition.
When the system becomes critical,  the central charge $c$  can easily be determined
from  the DMRG entanglement entropy.
Utilizing~Eq.~\eqref{se-cc}, Fig.~\ref{cc-gap}(c) demonstrates that $c^\ast$  indeed follows 
a linear fit to the DMRG data at the critical point (for $140\leq \chi\leq 400$), 
provided that prior to that the transition point was determined with extremely high precision.  
At $U/t=4$ and $J/t=-1.5$,  we have $c^\ast\simeq 0.499(1)$, suggesting the system to be in the Ising universality class where $c=1/2$.
For $U/t=4$ and $J/t=-0.5$ (not shown), we get $c^\ast\simeq 0.496(3)$. From conformal field theory~\cite{CC04}  the von Neumann entropy for a system with periodic boundary conditions takes the form $S_L(\ell)=(c/3)\ln\{(L/\pi)\sin[(\pi\ell/L)]\} +s_1$ with another  non-universal constant $s_1$. With a view of the doubled unit cell of the HI phase we slightly modify the related formula for $c^\ast$~\cite{Ni11}: 
\begin{eqnarray}
 c^\ast(L) \equiv \frac{3[S_L(L/2-2)-S_L(L/2)]}{\ln\{\cos[\pi/(L/2)]\}}\;.
\label{cstar}
\end{eqnarray}
Figure~\ref{susy} displays $c^\ast(L)$ when moving along the HI-CDW transition line by varying $U$ and $V$ simultaneously. 
Remarkably, when $U$ is raised, we find evidence for a crossover from  $c^\ast(L)\simeq 1/2$ to  $c^\ast(L)\simeq 7/10$,  which can be 
taken as a sign for an emergent supersymmetry at the boundary of the SPT HI phase~\cite{GSV14,FQS84,FQS85}.

Finally, we analyze the dynamical charge (spin) structure factor  of the model~\eqref{hamil},
\begin{equation}
 S^{(zz)}(k,\omega)=\sum_n |\langle\psi_n|\hat{\cal O}_k|\psi_0\rangle|^2\delta(\omega-\omega_n)\,,
 \label{duncor}
\end{equation} 
where $\hat{\cal O}_k=\hat{n}_k$ ($\hat{\cal O}_k=\hat{S^z}$). 
In Eq.~\eqref{duncor}, $|\psi_0\rangle$ ($|\psi_n\rangle$) denotes the ground ($n$th excited) state, and 
$\omega_n=E_n-E_0$.  Following Ref.~\cite{PVM12}, we first evaluate  the related two-point correlation functions,  
 $\langle \psi_0 |\hat{\cal O}_{j}(\tau) \hat{\cal O}_0(0) |\psi_0\rangle$, 
by way of real-time evolution of the ground-state infinite MPS
$|\psi_0\rangle$. Thereby we apply infinite boundary conditions
to a finite window of sites ($L_{\rm W}=128$). After $\hat{\cal O}$ 
is applied to a given site the system is evolved at least up to 
$\tau=30/t$, where a time step $\delta\tau=0.05/t$ is used in the
fourth-order Suzuki--Trotter decomposition.  Fourier transformation  then
gives the dynamical structure factors.

\begin{figure}[tb]
    \includegraphics[clip,width=0.9\columnwidth]{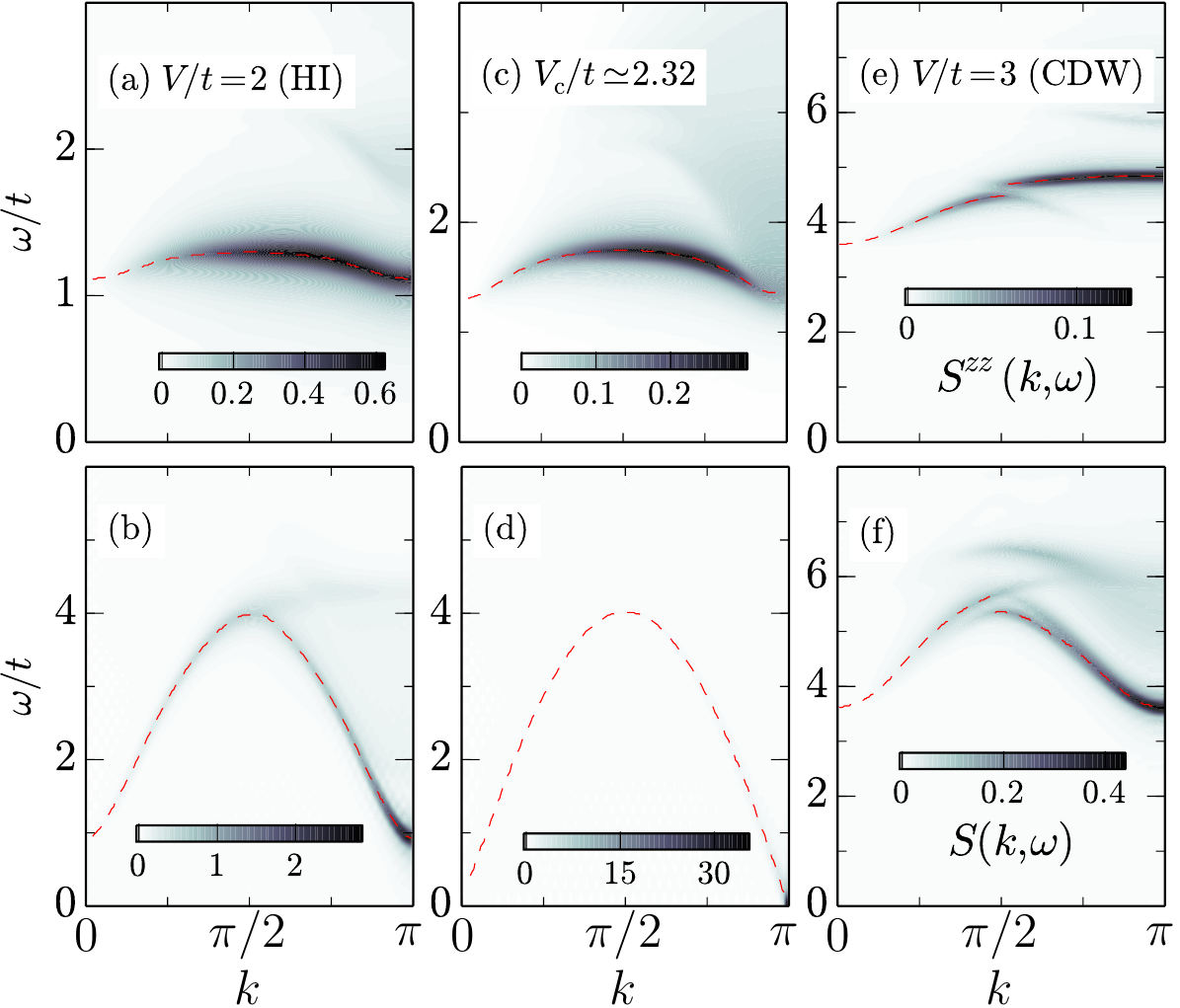}
   \caption{(Color online)
 Intensity plots of the dynamical spin structure factor $S^{zz}(k,\omega)$ (top) and density  structure factor
  $S(k,\omega)$  (bottom)  in the SPT HI phase (left), at the HI-CDW
 transition point (middle), and in the CDW phase (right). Dashed lines connect the intensity maxima at given $k$.
 Other model parameters are $U/t=4$ and $J/t= -1.5$.
 }
 \label{Skw-U4}
\end{figure}

For the spin-1 chain and extended Bose-Hubbard models it has been demonstrated that the dynamical spin
and density structure factor reveal distinguishing features in the SPT and topologically trivial  phases~\cite{EF15,ELF14,EF15b}.
Figure~\ref{Skw-U4} illustrates the intensity of the dynamical wave-vector-resolved spin and density response in the $k$-$\omega$
plane. In the HI phase, both  $S^{zz}(k,\omega)$ and $S(k,\omega)$
exhibit an essentially symmetric line shape with 
respect to $k=\pi/2$ and gaps at $k=0$ and $\pi$, but the spectral weight of the excitations is higher for $k>\pi/2$;  see Figs.~\ref{Skw-U4}(a) and \ref{Skw-U4}(b). 
While the spin response remains unaffected at the Ising transition point [Fig.~\ref{Skw-U4}(c)], the gaps in the charge response 
closes at $k=\pi$, reflecting the doubling of the lattice period  CDW phase [Fig.~\ref{Skw-U4}(d)]. 
Obviously $S(k,\omega)$ follows the behavior of the neutral gap rather than those of the charge gap [cf.  Fig.~\ref{cc-gap}(b)].
In the CDW phase,  the overall lineshape of  $S^{zz}(k,\omega)$ is asymmetric with a larger excitation gap at $k=\pi$. Note
that we find now two dispersive features (branches)  in  $S^{zz}(k,\omega)$ and $S(k,\omega)$, where a changeover of the intensity maximum 
takes place at $k=\pi/2$  [cf.  Fig.~\ref{cc-gap}(e) and (f)].

To summarize, exploiting the link between topological order and entanglement properties, we examined
the ground-state and spectral properties of the paradigmatic one-dimensional extended Hubbard model (EHM) with alternating
ferromagnetic spin coupling $J$ by numerically exact (DMRG) techniques. We showed that any finite spin interaction $J<0$ stabilizes  
a symmetry-protected topological Haldane insulator (SPT HI) that
replaces the spin-density-wave and bond-order-wave ground states existing in 
the pure EHM below a critical ratio of nearest-neighbor $(V)$ to intrasite $(U)$ Coulomb interaction. The HI manifests
the twofold degeneracy of the lowest entanglement level and, regarding the dynamical
 spin/density response,  reveals a similar  behavior as the SPT state of the spin-1 chain~\cite{EF15} and 
 the HI of the extended Bose-Hubbard model~\cite{ELF14,EF15b}.  Furthermore, analyzing the correlation length, entanglement spectrum and many-body excitation gaps, we found clear evidence for a quantum phase transition from the SPT HI phase to a CDW when the $V/U$-ratio is raised.  Using iDMRG, the HI-CDW boundary  and therefore the complete ground-state phase diagram could  be determined with very high accuracy.     
In the weak-to-intermediate interaction regime, the HI-CDW  transition belongs to the Ising universality class. Here the central charge $c=1/2$, and only the neutral gap vanishes. This is reflected in the dynamical density structure factor, where the gap closes at momentum $k=\pi$, just 
as for the HI-antiferromagnet transition of the spin-1 chain. In the strong interaction regime we found a first-order phase transition characterized by a jump in the spin gap. Decreasing the magnitude of $J$, the HI-CDW phase boundary approaches the BOW-CDW transition line in the pure EHM; thus, making the system topological, this changeover can be determined more precisely.  
Perhaps most interesting, tracing the central charge along the HI-CDW transition line, we detect a tricritical Ising point with $c=7/10$ that separates the continuous and first-order transition regimes. 
A further field theoretical study would be highly desirable to elucidate the origin of the tricritical Ising point. In either case the EHM with additional ferromagnetic spin exchange provides valuable insights into the criticality and nontrivial topological excitations of low-dimensional correlated electron systems. Note that we applied the ferromagnetic spin exchange in
order to easily realize an effective spin-1 state.
Including a physically more relevant dimerization of the transfer intergrals (hopping) 
will also stabilize the HI phase, so that the Ising quantum phase transition
occurs between the HI and CDW phases~\cite{BEG06}. Then, in this extended
Peierls-Hubbard model, the tricritical Ising point with $c=7/10$ will separate the HI-CDW transition 
line into continuous and  first-order lines ~\cite{Ejimaetal}.

{\it Note added in proof.} 
Due to the quantum-classical correspondence $D$-dimensional quantum and 
($D+1$) dimensional classical systems share important physical properties. 
So it is well known that the quantum spin-1 chain is related to  the classical two-dimensional restricted-solid-on solid (RSOS) model~\cite{RN87}.  It has been shown that a Fibonacci anyonic chain can be mapped-using the RSOS representation of the algebra-onto the tricritical Ising model with $c=7/10$. The transitions observed in our model can be understood as transitions from a low-density phase to a high-density phase of doubly occupied sites. Interestingly in the hard squares model first and second order transitions from low to high densities also occur with a tricritical point with the same central charge $c=7/10$, see~Ref.~\cite{KP91} and references therein. This connects-at the tricritical point-our model, hard squares and the so-called golden chain~\cite{FTLTKWF07}.

The authors would like to thank F.~G\"ohmann, T.~Kaneko, and  A.~Kl\"umper for valuable discussions. 
The iDMRG simulations were performed using the 
ITensor library~\cite{ITensor}. This work was supported by Deutsche Forschungsgemeinschaft (Germany), 
SFB 652.

\bibliography{ref}
\bibliographystyle{apsrev4-1}

\end{document}